\documentclass[a4paper,11pt]{article}
\pdfoutput=1 
\usepackage{jinstpub} 

\usepackage{float}
\usepackage{array}

\usepackage{etoolbox}
\usepackage{setspace}
\newcounter{rowcntr}[table]
\renewcommand{\therowcntr}{\thetable.\arabic{rowcntr}}

\newcolumntype{N}{>{\refstepcounter{rowcntr}\therowcntr}c}

\AtBeginEnvironment{tabular}{\setcounter{rowcntr}{0}}

\title{Studies on charging-up of single Gas Electron Multiplier}

\author[a,b,1]{Vishal Kumar,\note{Corresponding author.}}
\author[a,b]{Sridhar Tripathy,}
\author[c]{Purba Bhattacharya,}
\author[a,b]{Supratik Mukhopadhyay,}
\author[a,b]{Nayana Majumdar,}
\author[a,b]{and Sandip Sarkar}

\affiliation[a]{Applied Nuclear Physics Division, Saha Institute of Nuclear Physics, Kolkata - 700064, India}
\affiliation[b]{Homi Bhabha National Institute, BARC Training School Complex, Anushaktinagar, Mumbai, Maharashtra400094, India}
\affiliation[c]{Department of Physics, University of Cagliari and INFN, Strada prov.le per Sestu, km 1.00, 09042 Monserrato(CA), Italy}

\emailAdd{vishal.kumar@saha.ac.in}

\abstract{Mechanisms of charging-up and charging-down of Gas Electron Multiplier (GEM) have been studied. Experimental investigations have been carried out on both dielectric polarization and radiation charging of GEMs. Environmental parameters, such as pressure and temperature have been monitored to normalize their effects on the charging-up and charging-down measurements. Variation in gain due to the combined, as well as individual, effects of the mentioned parameters, have been illustrated.}

\keywords{Single GEM detector, charging-up, charging-down, gain, radiation rate, polarization.}

\begin{document}
\maketitle
\flushbottom

\section{Introduction}
\label{S1}

Many of the present gaseous ionization detectors have exposed dielectric components within the gas volume. When the gas gets ionized and avalanches, or streamers, occur within the volume, there is a possibility that the resulting electrons and ions come in contact with the dielectric material. Some of these charged particles get attached to the material, depending on several factors such as the presence of an impurity, defects of the material itself, and the kinetic properties of the impinging particles \cite{Rau2019}. This topic is rather complex and an area of study by itself because the dynamics affect several processes related to both academic and practical importance.

Study of the dynamics of charging-up and its effects are also important for design and optimum utilization of gaseous detectors. This is especially true for devices such as the Gas Electron Multiplier (GEM) \cite{sauli2016gas} that has a large amount of insulating material exposed to the active gas volume. Usually, a simple dynamical model is assumed to interpret the process and effects of charging-up in these devices. In such simple terms, the charging-up phenomenon in GEM takes place mainly due to two reasons, namely, the polarization of dielectric in a high electric field and the accumulation of charges on the dielectric during amplification. In both cases, the charge configuration inside the GEM hole changes leading to a modified field configuration. This, in turn, affects the detector response.

The charging-up phenomenon in GEM has been observed way back in 1997 \cite{bouclier1997new} and later described in detail by C. Altunbas \cite{altunbas2002construction}.
While use of slightly conducting substances (such as water, or diamond-like carbon coating) has been proposed as remedies to the gain shift due to charging-up \cite{han2002,azmoun2006}, these investigations identified several factors (geometry of the foil, rate of ionizing radiation, water content in the operating gas) that contribute towards the gain stability related to the charging-up process. In \cite{croci2009}, detailed study of the charging-up process has been reported with conclusions that are seemingly in contrast to \cite{han2002,azmoun2006}. In recent times, there has been a flurry of activities on this problem, probably because of its importance in the high rate experiments that scientists are embarking upon. Several numerical \cite{alfonsi2012simulation,correira2014} and experimental \cite{hauer2019} efforts have tried to resolve the problem by careful analysis.

In this work, we have made attempts to contribute to the resolution of the charging-up problem by carrying out detailed experimental studies.
Since, according to several interpretations \cite{croci2009,alfonsi2012simulation,correira2014}, the charging-up process includes two sub-processes (polarization of dielectric and accumulation of charges on the dielectric), we have tried to identify the effects of these two processes on the effective gain stability, separately, as well as in combination.

The standard GEM foil of 10 cm by 10 cm procured from CERN has been used in the experimental measurements. The GEM foil is made up of a 50 $\mu$m polyimide sheet sandwiched between two copper layers of 5 $\mu$m, with biconical holes with inner and outer diameters as 50 $\mu$m and 70 $\mu$m respectively. These biconical holes are etched out by chemical lithographic technique in a hexagonal pattern with 140 $\mu$m pitch.

Detector systems based on similar GEM foils are used in a large number of applications, including in-beam high energy experiments. Due to its excellent rate handling capability \cite{abbaneo2015}, GEM-based detectors are used regularly for high luminosity experiments \cite{bellini,abbaneo2014}. Because of its material properties, the issues of charging-up and charging-down of GEM foils are invariably present in many of these applications. According to our observations, the gain of the GEM-based detector depends significantly on both the rate and duration of radiation. Thus, especially in those experiments where beam current changes appreciably with time, our observations could play an important role in calibrating gain of these detectors.

\section{Experimental Setup} \label{S2}
\begin{figure}[htbp]
\centering
\includegraphics[width= 0.55\linewidth ,keepaspectratio]{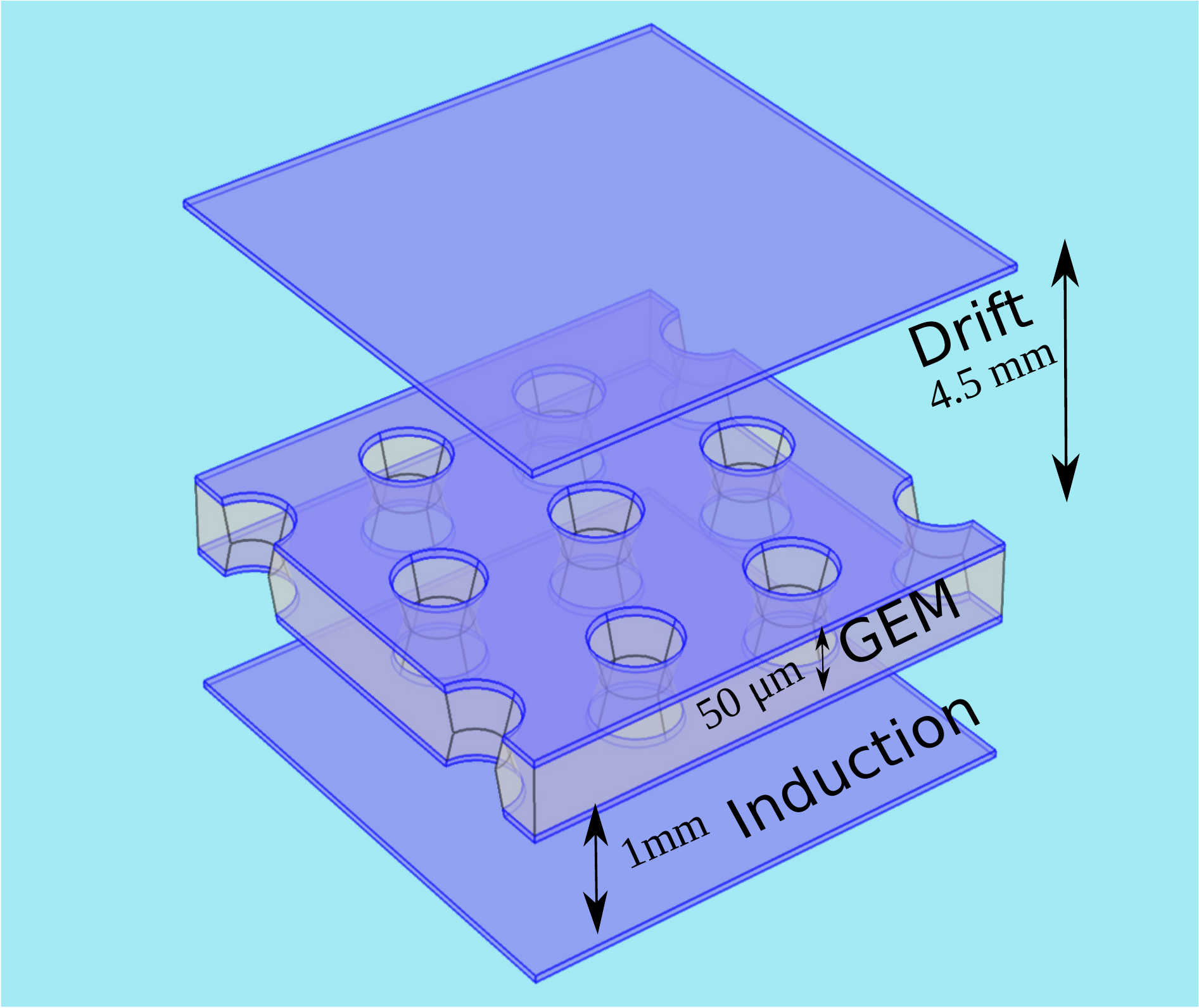}
\caption{\label{fig1} Schematic diagram of single GEM detector.}
\end{figure}

The experimental setup consists of a single GEM foil mounted between a cathode and an anode plane with a drift gap of 4.5 mm and an induction gap of 1 mm as shown in figure~\ref{fig1}. The anode is comprised of 256 readout strips grouped into two, containing 128 strips each, in both x and y-sensing planes and are grounded using 120 k$\Omega$ resistors. The combined strips divide the GEM active area into four sectors with two x and two y-sensing plane signal outputs.

\begin{figure}[htbp]
\centering
\includegraphics[width= 0.65\linewidth,keepaspectratio]{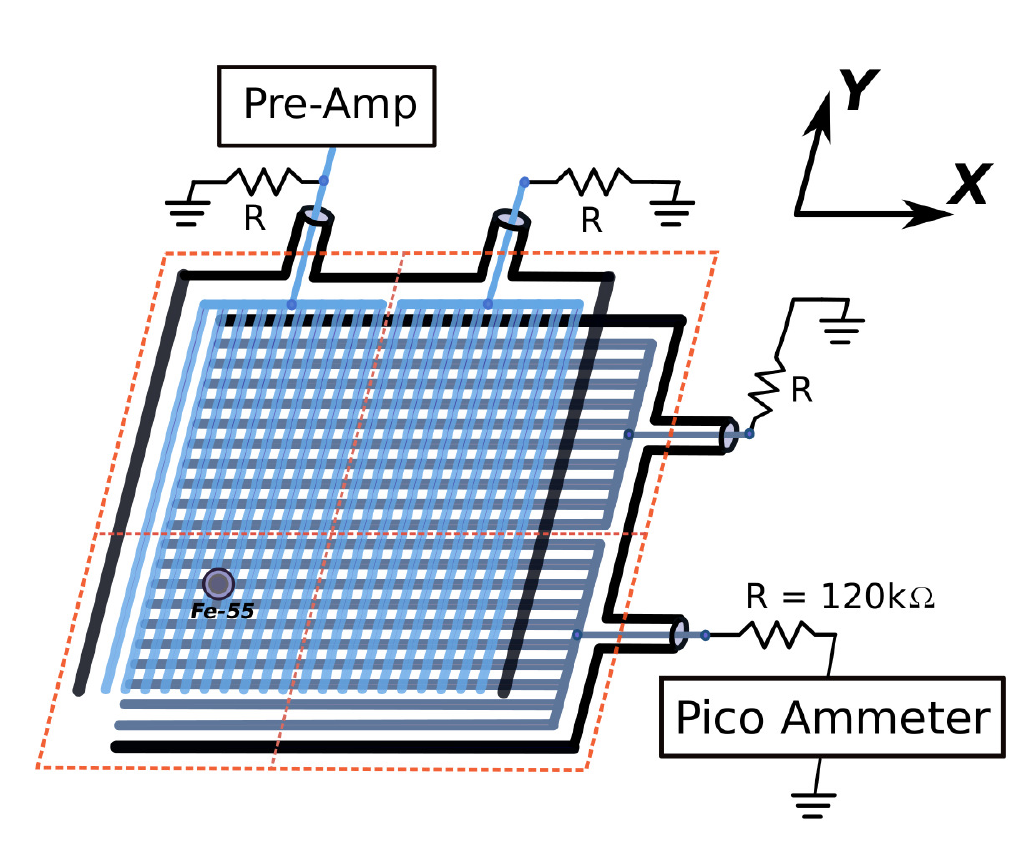}
\caption{\label{fig2} Schematic diagram of GEM readout and its connections.}
\end{figure}

The measurements have been done by connecting a section of the x-sensing plane to a CAEN A1422 preamplifier \cite{caenA1422} and CAEN AH401D picoammeter \cite{caenAH401D} to the y-sensing plane for energy and current measurements, respectively, as shown in figure~\ref{fig2}. Alongside these measurements, the temperature, pressure and humidity have been constantly monitored and stored with an Arduino Uno R3 \cite{UNO} using a Bosch BME280 sensor \cite{Bosch}.

The drift, induction and amplification fields have been achieved by applying potential differences, as described in table~\ref{table1}, with some modifications according to the experimental requirements. These field configurations have been optimized to get maximum gain allowing a negligible electrical discharge probability in Ar-CO$_2$ (74-26 $\pm 1\%$) gas mixture, employed in all measurements, except those reported in section 8. In the mentioned section, the gas mixture is Ar-CO2 at 90:10 volumetric ratio.

\begin{table*}[htbp]
\begin{center}
\caption{Experimental configuration} \label{table}
\scalebox{0.83}{
\begin{tabular}{ N c l l l l l l }
\hline
\multicolumn{1}{c}{Serial}& Description & E$_{Drift}$ & $\Delta$V$_{GEM}$ & E$_{Ind.}$ & Aperture & Equivalent \\
\multicolumn{1}{c}{No.}& & (kV/cm) & (Volts) & (kV/cm) & size (mm) & Rate (kHz) \\
\hline
\label{i} & Gain measurement& 3.44 & 508.35 & 4.99 & 6.0 & 4.0 $\pm~2.3\%$ \\
\label{ii} & T/P correction &3.44 & 508.35 & 4.99 & 2.0 & 0.49 $\pm~2.3\%$ \\
\label{iii} & $\Delta$V$_{GEM}$ variation & 3.44 & 462.15 - 508.35 & 4.99 & 6.0 & 4.0 $\pm~2.3\%$ \\
\label{iv}&Eq. rate variation& 3.44 & 508.35 & 4.99 & 1.0 - 10.0 & 0.12 - 25.5 $\pm <5.0\%$ \\
\label{v}& Eq. rate variation & 3.44 & 508.35 & 4.99 & 4.0 - 10.0 & 1.4 - 25.5 $\pm <5.0\%$ \\
\label{vi} & $\Delta$V$_{GEM}$ variation& 3.44 & 462.25 - 518.80 & 4.99 & 6.0 & 4.0 $\pm~2.3\%$ \\
\label{vii}&Charging-down& 3.44 & 508.35 & 4.99 & 2.0 & 0.49 $\pm~2.3\%$ \\
\hline
\end{tabular}}
\label{table1}
\end{center}
\end{table*}

The measurements have been done using an $^{55}$Fe extended source of diameter 1.2 cm whose rate of radiation has been controlled by collimating the source using collimators having different apertures. The collimator is a disk made up of stainless steel of thickness 3 mm with an outer diameter of 5 cm having a circular aperture at its centre with the diameter ranging from 1 to 10 mm. Moreover, to obtain experimental data on polarization charging-up (section \ref{S5}), it has been important to use a highly collimated (1 mm in diameter) radiation source so that the effect of radiation on charging up / down is minimized during these measurements.

\subsection{Equivalent rate estimation} \label{S2a}
For a given collimator, a large number of spectra have been obtained by repeated measurements and fitted 5.9 keV x-ray spectra by Gaussian. The area under such a gaussian has been considered to be the total number of counts for that measurement. Finally, the equivalent rate of radiation for each measurement has been estimated by dividing the count by the duration of that measurement.

\begin{itemize}
\item The equivalent rates so calculated represent the amount of 5.9 keV x-ray getting detected after entering the detector through the collimator.
\end{itemize}

\begin{figure}[htbp]
\centering
\includegraphics[width= 0.65\linewidth,keepaspectratio]{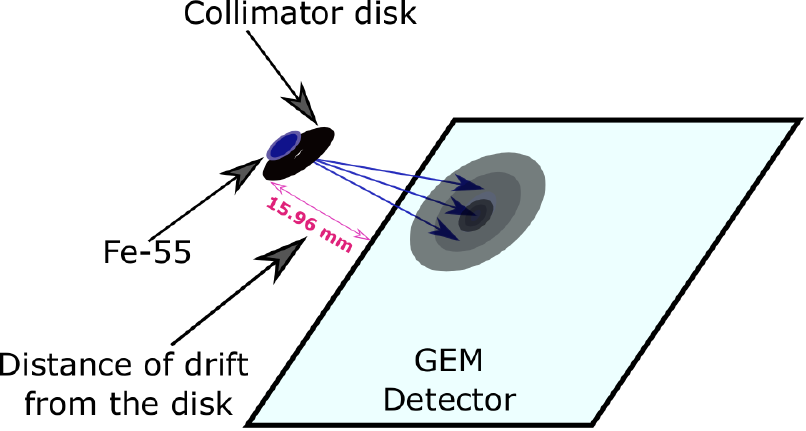}
\caption{\label{fig2a} Schematic diagram of radiation rate profile on the detector after passing through the collimator.}
\end{figure}

The collimator as mentioned in section \ref{S2} is a thin disk and has been used to control the equivalent rate of x-ray falling on the detector. The radiation beam is not parallel and makes a Gaussian profile on the detector as shown in figure~\ref{fig2a}. This fact has been verified both experimentally and numerically.

\begin{figure}[htbp]
\centering
\includegraphics[width= 0.65\linewidth,keepaspectratio]{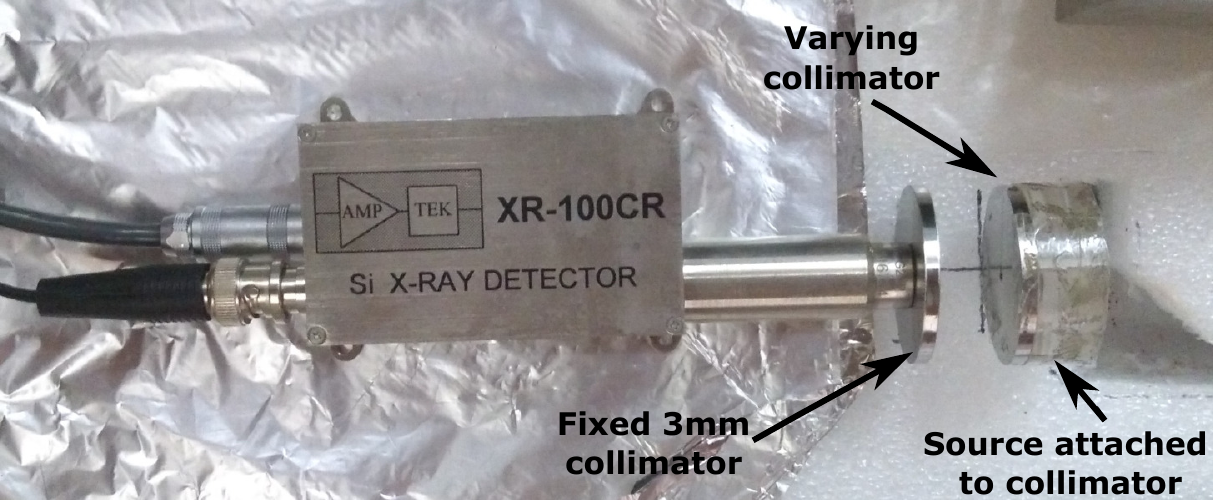}
\caption{\label{fig2b} Experimental setup with Si PIN Photodiode detector.}
\end{figure}

\begin{figure}[htbp]
\centering
\includegraphics[width= 0.65\linewidth,keepaspectratio]{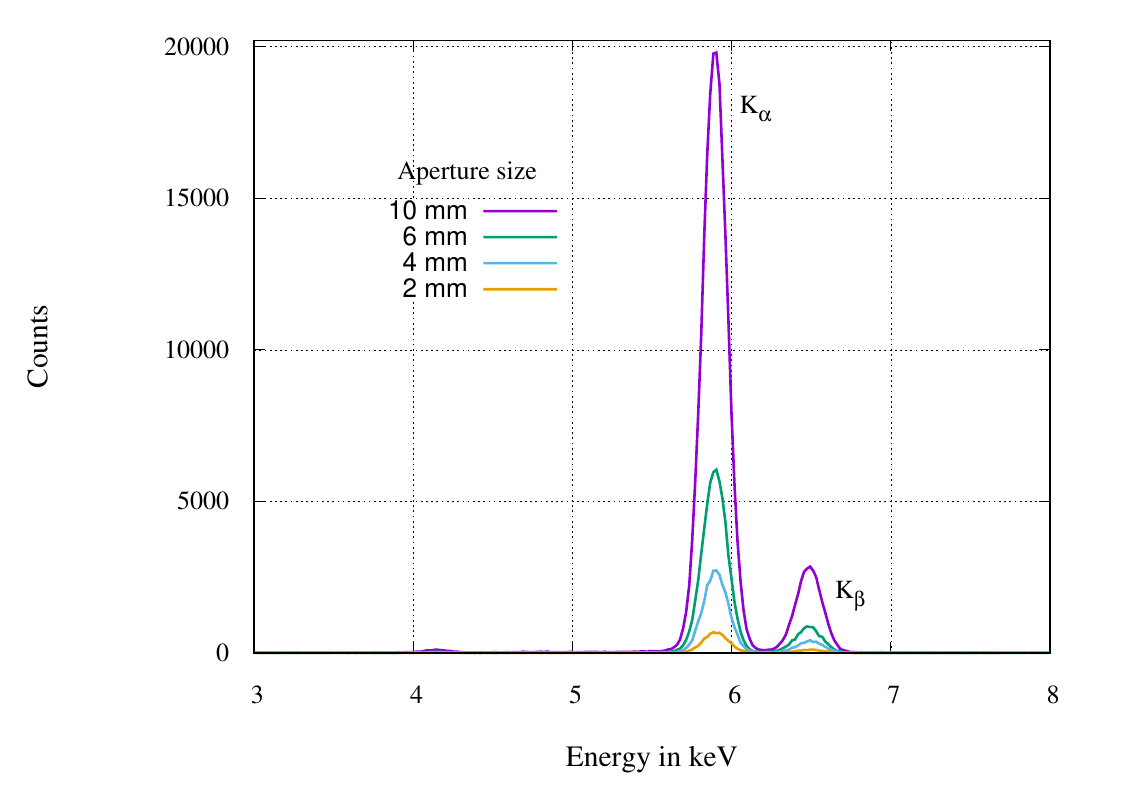}
\caption{\label{fig2b1} Energy spectra obtained by Si PIN Photodiode from collimators with different aperture sizes.}
\end{figure}

\begin{figure}[htbp]
\centering
\includegraphics[width= 0.65\linewidth,keepaspectratio]{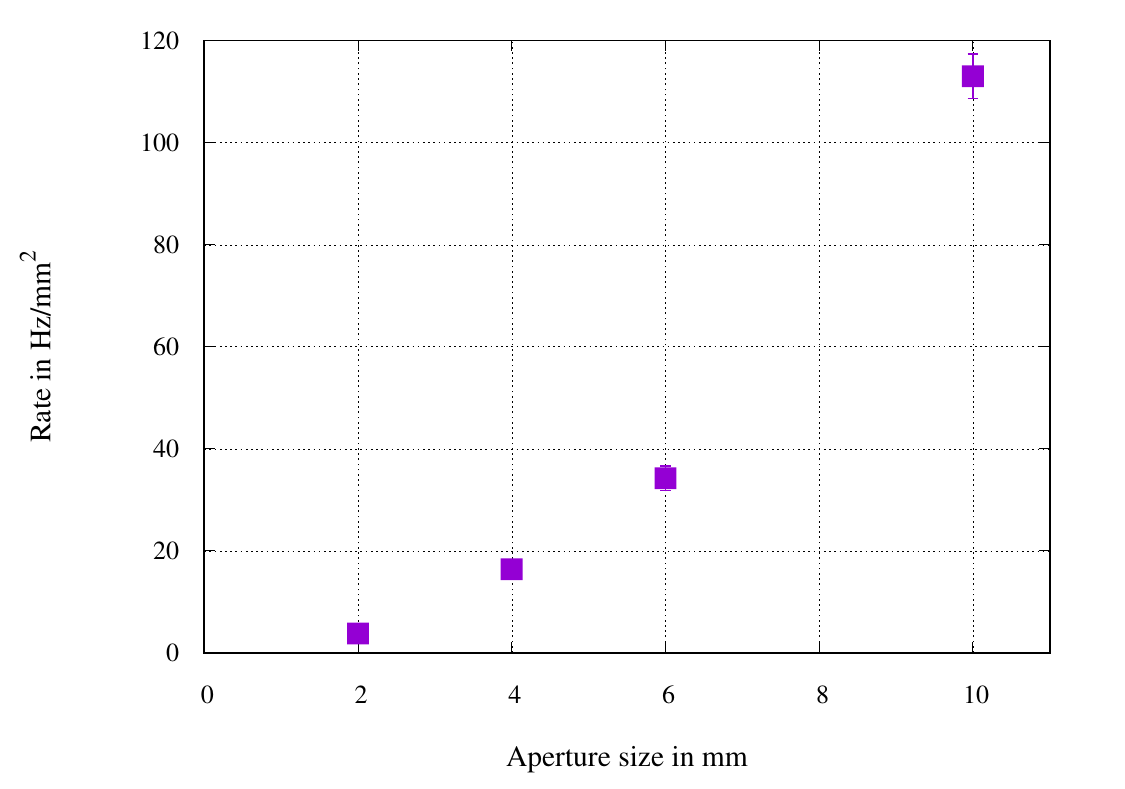}
\caption{\label{fig2c} Radiation rates of 5.9 keV x-ray detected by Si PIN Photodiode with aperture size of collimators.}
\end{figure}

In order to verify it experimentally, a similar setup of source and disk have been used only this time the GEM detector has been replaced by a Silicon PIN Photodiode detector  with sensitive area of 6 mm$^2$ \cite{PIN} and a collimator disk with 3 mm aperture size placed in front of it as shown in figure~\ref{fig2b}. The distance between the photodiode detector and the collimator attached to the source is fixed at 15.96 mm which is also the distance between the collimator and drift plane in case of GEM detector. The experiment has been carried out by recording energy spectra as shown in figure~\ref{fig2b1} by varying the aperture size of the collimator disk attached to the radiation source. Figure~\ref{fig2c} shows the variation of the rate per unit time per unit area with aperture size. The collimator fixed to the Si PIN diode ensures that the area of detection remains constant making it clear that the rate have been modified with the help of collimator placed in front of the extended radiation source.

\begin{figure}[htbp]
\centering
\includegraphics[width= 0.55\linewidth,keepaspectratio]{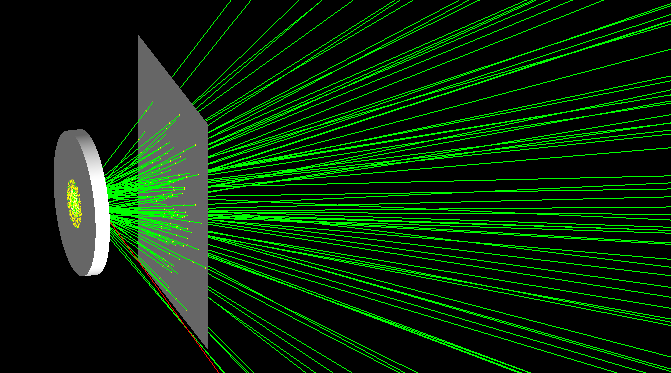}
\caption{\label{fig2d} GEANT4 geometry used for simulation.}
\end{figure}

\begin{figure}[htbp]
\centering
\includegraphics[width= 0.5\linewidth,keepaspectratio]{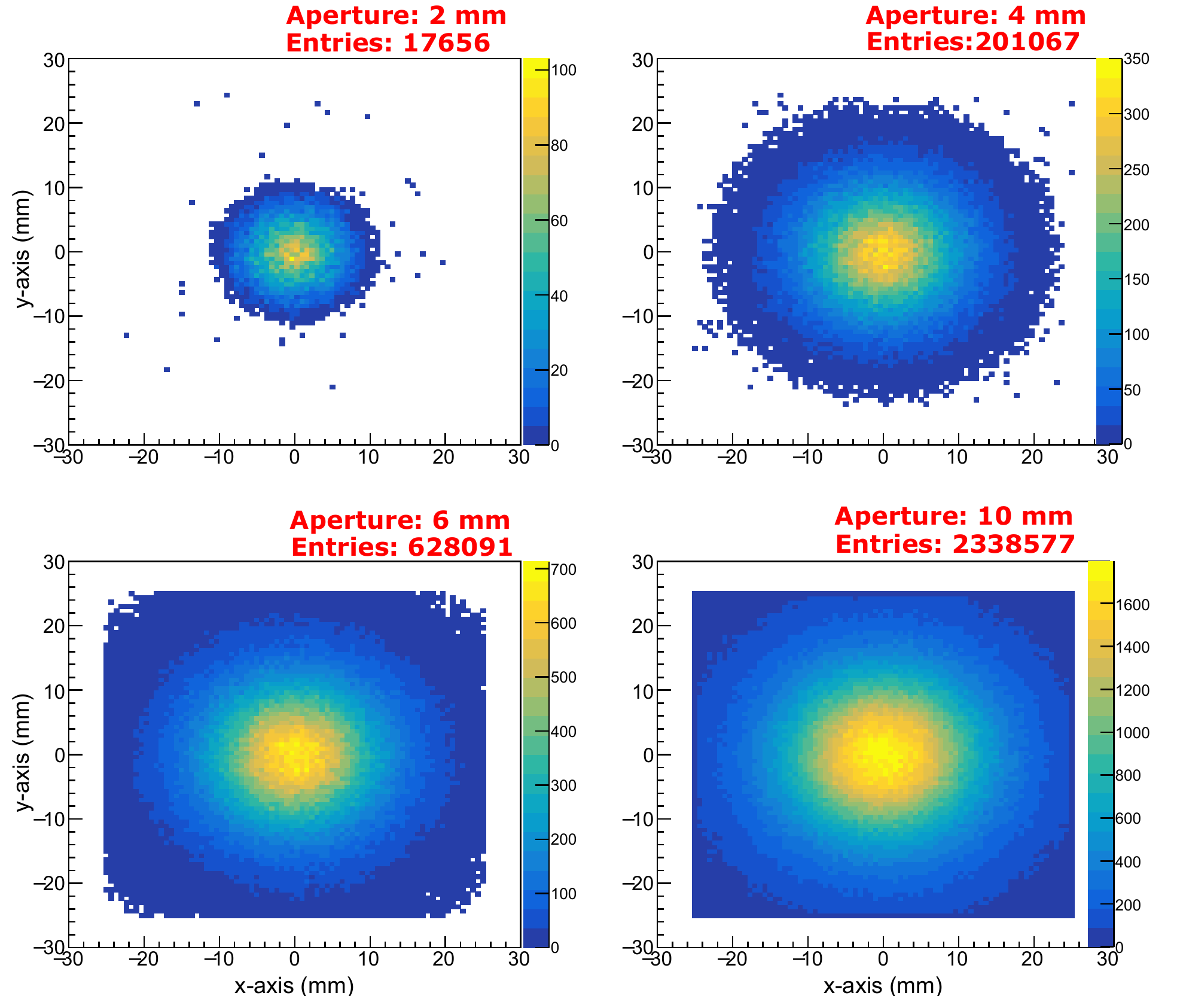}
\caption{\label{fig2e} Event profile on 25 mm by 25 mm screen placed at 15.96 mm away from collimator generated by GEANT4 to model the rate of single GEM experiment.}
\end{figure}

Numerically, the rate variation due to change in collimator aperture size is been studied with the help of GEANT4 \cite{GEANT} simulation. The geometry of this model is similar where a collimator disk is placed in front of an extended source producing 5.9 keV x-ray. A total of $5 \times 10^6$ events of x-rays were incident on the sensitive plane through the opening of the collimator uniformly in the forward hemisphere as shown in figure~\ref{fig2d}. The aperture size of the collimator is varied and the 3D Gaussian profile so obtained is shown in figure~\ref{fig2e}.

\begin{figure}[htbp]
\centering
\includegraphics[width= 0.5\linewidth,keepaspectratio]{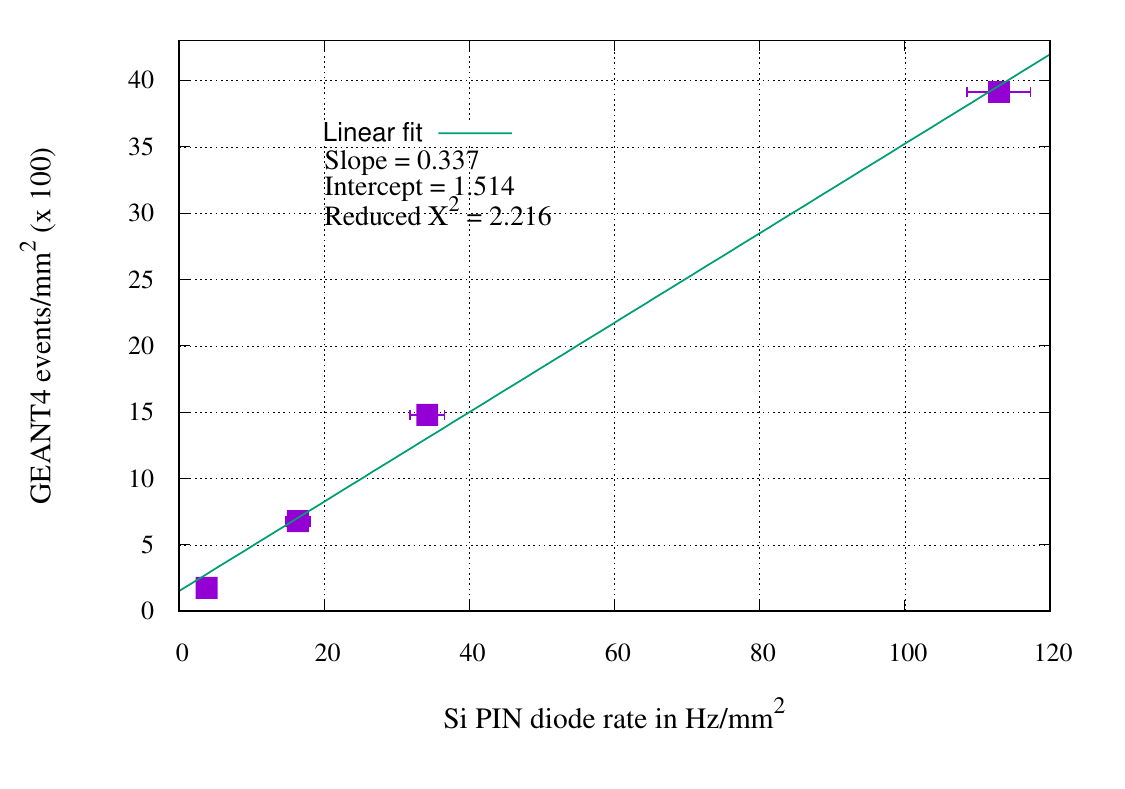}
\caption{\label{fig2f} Comparison of the rate observed using Si PIN diode detector and events rate generated by GEANT4 inside  the circular region of 6 mm$^2$.}
\end{figure}

To compare both the results obtained from the experiment using Si PIN diode and numerical simulation using GEANT4. The total number of hitpoints within a circular area of 6 mm$^2$ (equal to sensitive area of Si PIN diode) , centred at (0,0) have been calculated. The rate and hitpoints so obtained were plotted in figure~\ref{fig2f}. These values are nearly linear, confirming the fact that the rate per unit area have changed by using thin collimator of different aperture size.

\section{Gain measurement}
\label{S3}
The ratio of the number of electrons collected by readout planes to the number of primaries generated in the drift region is called effective gain. It mainly depends upon the amount of multiplication of primaries in the amplification region along with collection and extraction efficiencies which are dependent on field variation at the interfaces between drift and amplification, and amplification and induction regions.

For effective gain measurement, the output signal from the preamplifier attached to the readout x-sensing plane has been amplified using ORTEC 672 spectroscopy
amplifier \cite{Ortec672}. The amplified Gaussian pulse has been fed to Amptek MCA-8000D \cite{MCA} digital multichannel analyser to generate the energy spectrum in a computer.
Simultaneously, the current has been measured from the picoammeter connected to the y-sensing plane readout. These measurements have been used to calculate the effective gain (G$_{eff}$) using the formula mentioned in \cite{patra2017} after some modifications, as given below:
\begin{equation}
\label{eq1}
G_{eff}=\frac{I\cdot \Delta t}{\sum_i{Q_i}}
\end{equation}
\begin{equation}
\label{eq2}
Q_i=N_i\cdot p_i\cdot e
\end{equation}
where, $\Delta$t is the time interval for collecting the energy spectra, I is average current from the picoammeter over the interval $\Delta$t , e is charge of an electron, N$_i$ is the number of counts in i$^{th}$ channel of the spectrum and p$_i$ is its corresponding number of primaries. The numerator $I\cdot\Delta t$ is total charge collected at the anode during the time interval $\Delta t$ and denominator $\sum_i{Q_i}$ is the total charge of the primaries created inside the detector during the same period. The gain measurement process has been repeated for each spectrum taken for $\Delta$t time.

\begin{figure}[htbp]
\centering
\includegraphics[width= 0.65\linewidth,keepaspectratio]{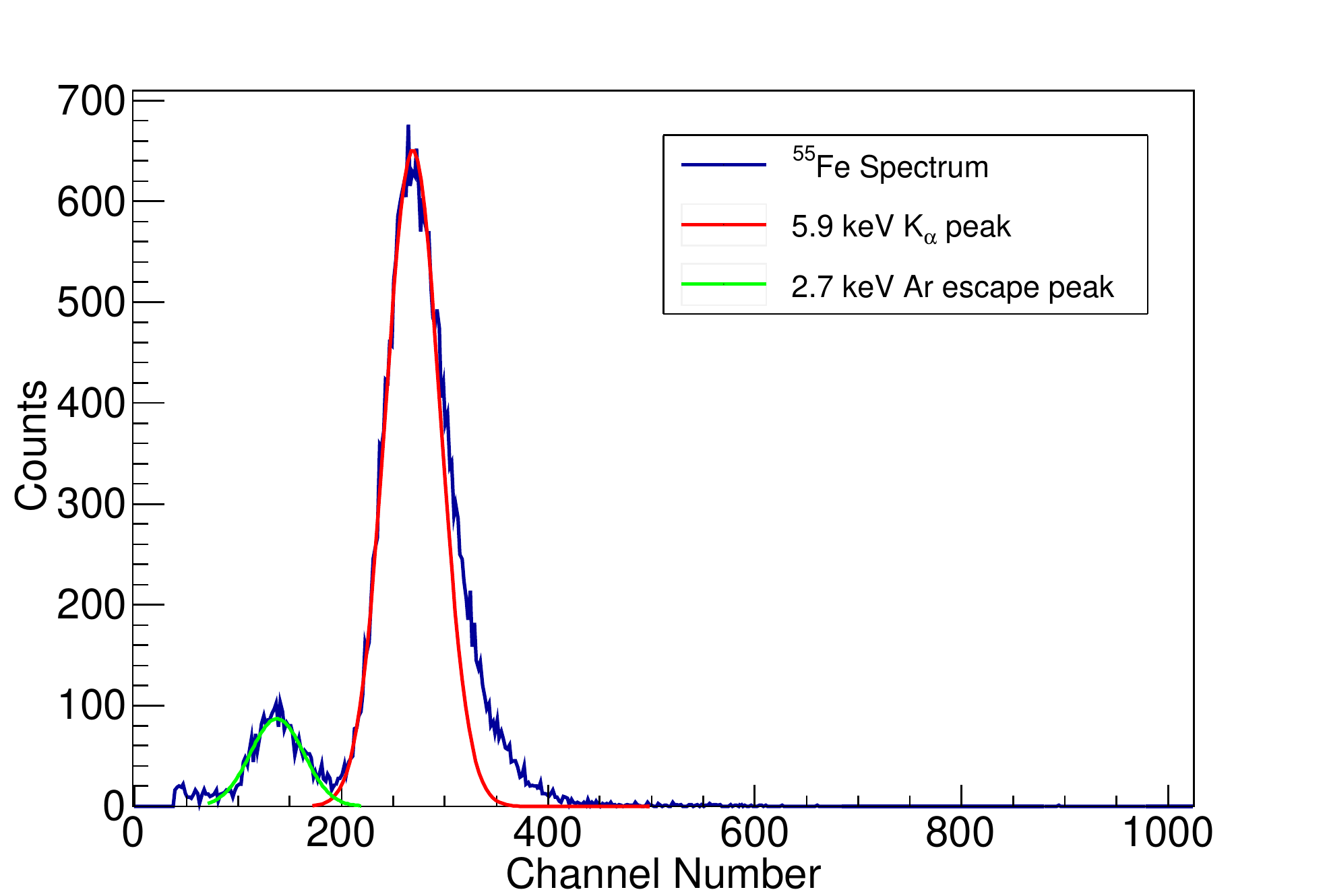}
\caption{\label{fig3} Peaks fitting in $^{55}$Fe spectrum used for gain measurement with configuration \ref{i}. The fitting range for K$_{\alpha}$ have been selected to avoid contributions from K$_{\beta}$ peak.}
\end{figure}

In order to calculate the primaries corresponding to i$^{th}$ channel of the spectrum, p$_i$, the MCA channels have been converted to energy values by two-point calibration by fitting $^{55}$Fe 5.9 keV x-ray and 2.7 keV Ar escape peaks with Gaussian as shown in figure~\ref{fig3}. These energy values have been further converted into the number of primaries with the help of data generated by Garfield++ using HEED \cite{HEEDv2.10,smirnov2005} for Ar-CO$_2$ (74-26\%) gas mixture to give p$_i$ values. By calculating the p$_i$ values by this method and using it in \eqref{eq2} gives a better estimate of primaries for gain calculation instead of using fixed number of primaries over the Gaussian. To increase the accuracy of gain measurement, the spectrum which was not fitted properly and have high sigma values were removed during the calculation.

The spectrum (figure~\ref{fig3}) has been generated using $^{55}$Fe with a 6 mm collimator giving a equivalent rate of 4.0 kHz producing a measurable current for the picoammeter. This rate, as discussed earlier (section \ref{S2a}), has been calculated by taking the area under the Gaussian fit of $^{55}$Fe 5.9 keV x-ray spectra and dividing it by the time interval $\Delta$t. The time interval $\Delta$t = 5 sec has been optimized to get sufficient counts in the spectrum for Gaussian fitting and at the same time small enough to capture the changes caused by charging-up with time. The $\Delta$t value has been modified for measurements at other radiation rates, as per requirement. The effective gain measured for each spectrum taken with an interval of $\Delta$t has been used to calibrate the centroid of 5.9 keV x-ray Gaussian and this calibration has been used to determine the gain for all the experiments. This is an effective method for gain measurement since the current measurement for low rate experiment is difficult and some times not possible.

\begin{figure}[htbp]
\centering
\includegraphics[width= 0.65\linewidth,keepaspectratio]{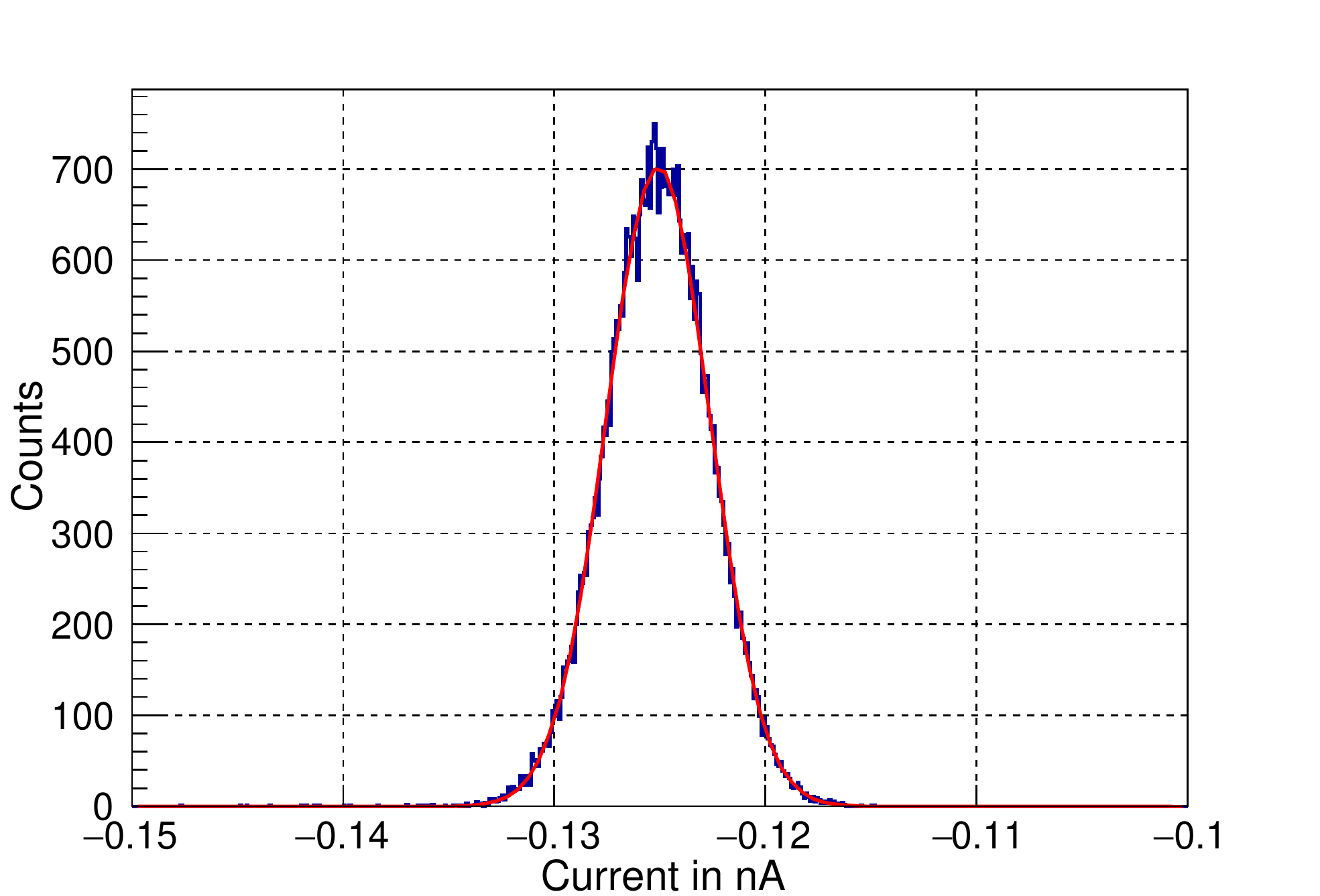}
\caption{\label{fig3b} Gaussian fit for repeated current measurement with 4.0 kHz $^{55}$Fe source to get the accuracy in the current with configuration \ref{i}.}
\end{figure}

The current has been measured using CAEN AH401D picoammeter with a capacitance value of 350 pC and integration time of 280 ms setting the range of current that can be measured upto 1.25 nA with an accuracy of 1.25 fA as mentioned in the manual \cite{caenAH401D}. However, the accuracy is also estimated experimentally by repeated current measurement with 4.0 kHz collimated $^{55}$Fe source. These current values are fitted with Gaussian as shown in figure~\ref{fig3b} with mean -0.125 nA and sigma 2.47 pA ($<2 \%$). It may be noted here that this measurement includes fluctuations due to temperature variation. The picoammeter settings remain the same throughout the experiment to ensure consistency.

\section{Effect of environmental parameters on gain}
\label{S4}
\begin{figure}[htbp]
\centering
\includegraphics[width= 0.65\linewidth,keepaspectratio]{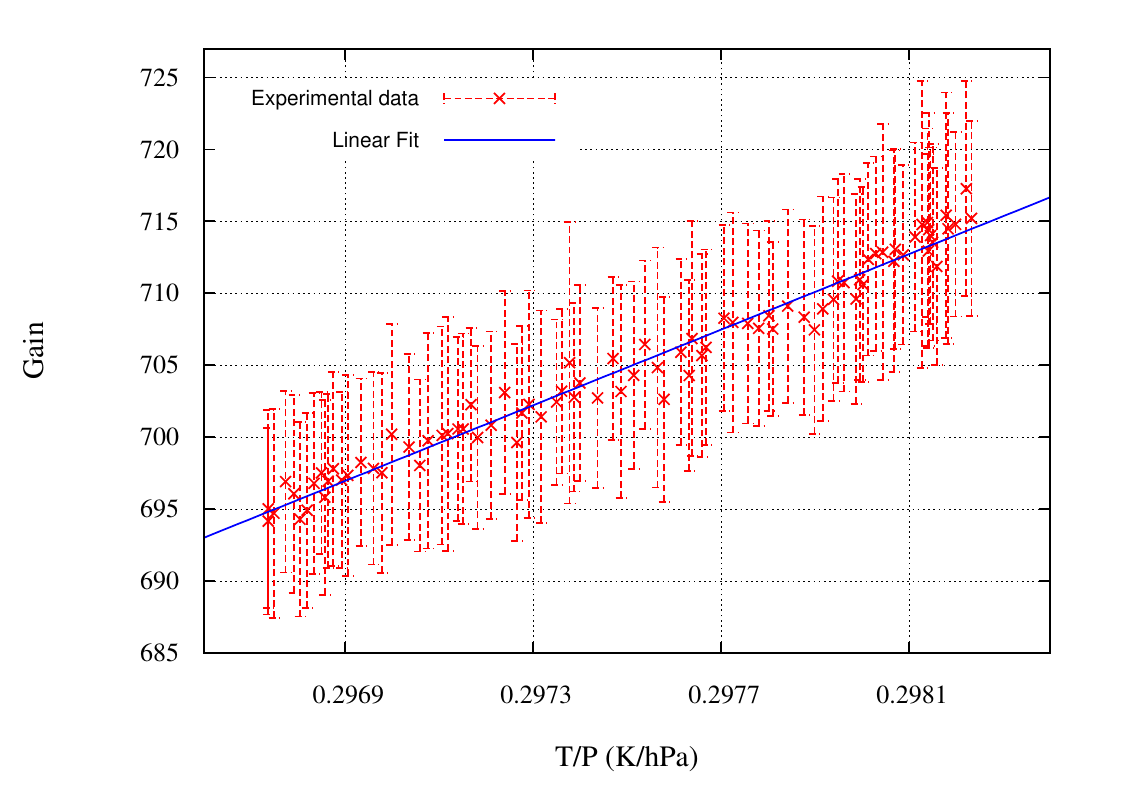}
\caption{\label{fig4} Linear dependence of GEM gain with T/P, with slope 13029.8 hPa/K at 490 Hz \ref{ii}.}
\end{figure}

\begin{figure}[htbp]
\centering
\includegraphics[width= 0.65\linewidth,keepaspectratio]{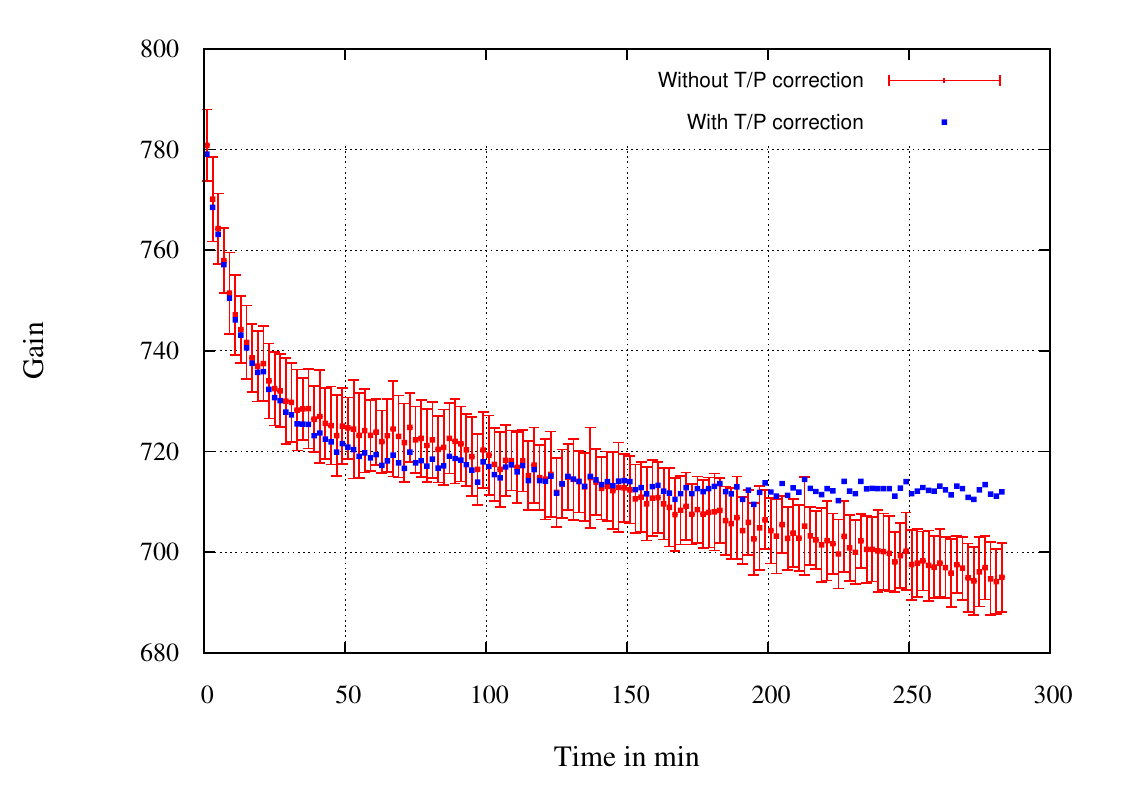}
\caption{\label{fig5} Temperature and pressure correction of radiation charging-up data with configuration \ref{ii}.}
\end{figure}

The gain is significantly affected by environmental parameters. It is exponentially dependent on the ratio of temperature and pressure (T/P) \cite{altunbas2003aging}. The experiment has been conducted at room temperature and pressure, with small natural variations in them during experimental measurements. In figure~\ref{fig4}, the gain versus T/P is plotted where the exponential behaviour has been approximated by a linear dependence since the x-axis range is very small.
These linear fit results have been used to normalize the effect of environmental parameters. The corrected data are plotted in figure~\ref{fig5} along with the uncorrected one. It shows that the effect is quite significant.

\section{Charging-up due to polarization of GEM dielectric}
\label{S5}

The GEM foil has a very high electric field (amplification field) across its dielectric. The dielectric gets polarized due to such a high field. This is known as charging up due to polarization of dielectric and, as a consequence, the gain increases.

\begin{figure}[htbp]
\centering
\includegraphics[width= 0.6\linewidth,keepaspectratio]{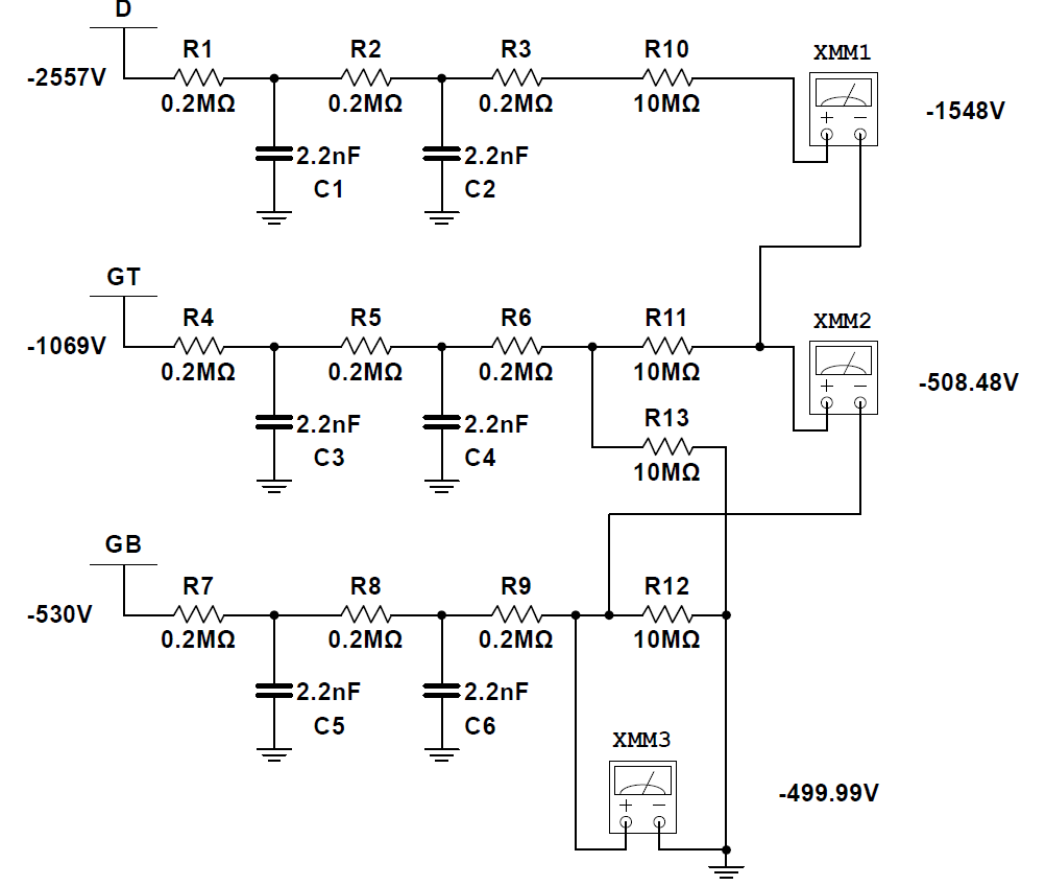}
\caption{\label{fig5b} High voltage filter circuit used to provide respective voltages to the detector.}
\end{figure}

To ensure that the detector dielectric is completely unpolarized, the following measure has been taken:
\begin{itemize}
\item The detector has been kept without bias and radiation for several days.
\end{itemize}
The CAEN N471 \cite{N471} voltage supply with a ramp-up rate of 200 V/s has been used to provide the necessary potential to the electrodes with the help of a potential divider circuit which has an in-built high voltage filter as shown in figure~\ref{fig5b}.

\begin{figure}[htbp]
\centering
\includegraphics[width= 0.65\linewidth,keepaspectratio]{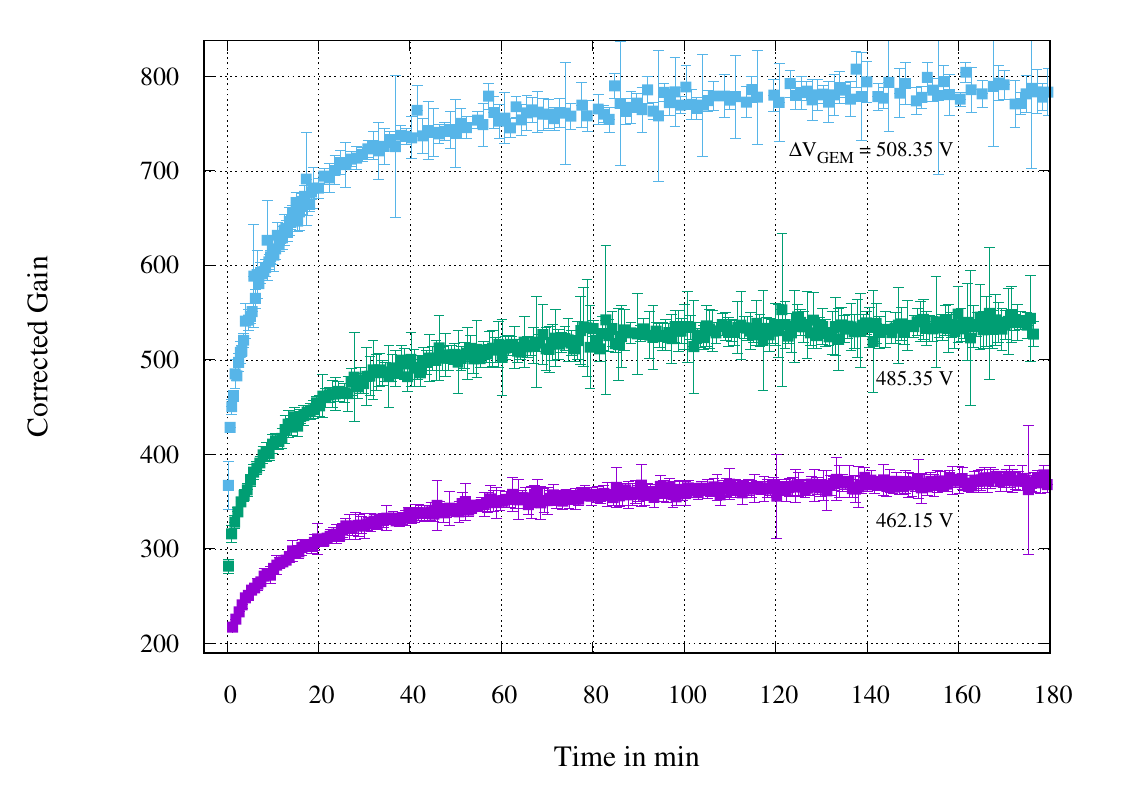}
\caption{\label{fig6} Polarization charging-up of dielectric at different GEM voltages with configuration \ref{iii}.}
\end{figure}

After the potential has been attained within 15 to 20 seconds (ramping-up at 200 V/s), the effect of dielectric polarization on the gain has been measured with the help of a very small test probe of 120 Hz rate (minimizing radiation charging-up effect). It has been studied with different GEM voltage values and the results are as shown in figure~\ref{fig6}. It is clear from the figure that the effect of polarization charging is to increase the gain of the detector. The effect is more prominent with the increase of applied voltage. Finally, the gain stabilizes to a saturation value after almost 3 hours of voltage application.

\begin{figure}[htbp]
\centering
\includegraphics[width= 0.65\linewidth,keepaspectratio]{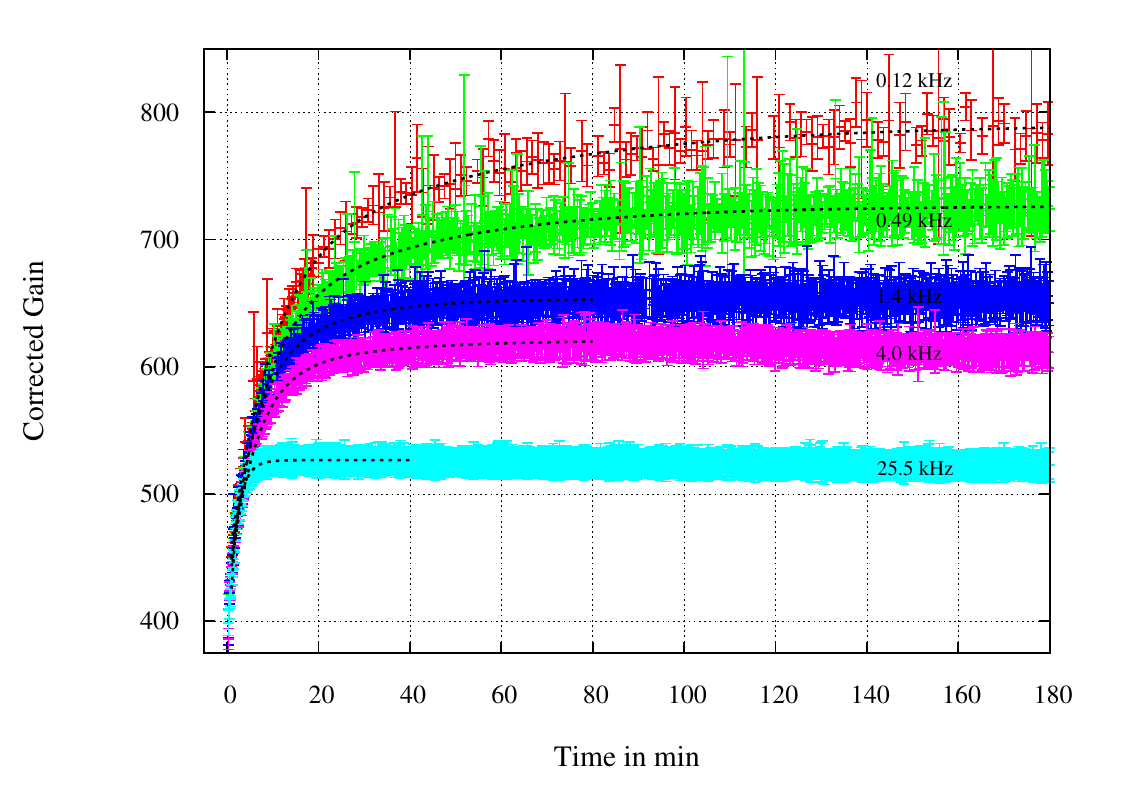}
\caption{\label{fig7} Combined effect dielectric polarization and radiation charging-up with configuration \ref{iv}.}
\end{figure}

The gain due to the application of $\Delta$V$_{GEM} = 508.35~V$, measured with different radiation rates are shown in figure~\ref{fig7}. These results are further fitted by function \ref{eq3} mentioned in \cite{Jonuz}. The fitting is done in the time range where polarization charging-up takes place in order to avoid unnecessary fluctuations due to other factors which occur after gain saturation, the parameters are as shown in table~\ref{table2}.
\begin{equation}
\label{eq3}
G_{eff}(t)=A (1- e^{\frac{-t}{a}}) + B (1-e^{\frac{-t}{b}})+C
\end{equation}
\begin{table*}[htbp]
\begin{center}
\caption{\label{table2} Fit Parameters}
\scalebox{1.0}{
\begin{tabular}{ l l l l l l l }
\hline \\
Eq. rate& Range & A & a & B & b&C \\
(kHz)& (min) & &(min) & &(min) & \\
\hline \\
0.120&	1-180	&	237.027&	8.776&	121.510&	48.884&432.435 \\
0.490&	1-180 &	208.370&	7.438&	96.919&	36.171&	421.090 \\
1.40&	1-80	&	72.008&	16.518&	166.081&	5.278&	415.266 \\
4.0& 1-80 &	34.127&	22.867&	173.901&	5.530&	412.764 \\
25.5&	1-40	&	80.169&	1.106&	112.289&	2.034&	334.060 \\
\hline
\end{tabular}}
\end{center}
\end{table*}

\begin{figure}[htbp]
\centering
\includegraphics[width= 0.5\linewidth,keepaspectratio, angle=-90]{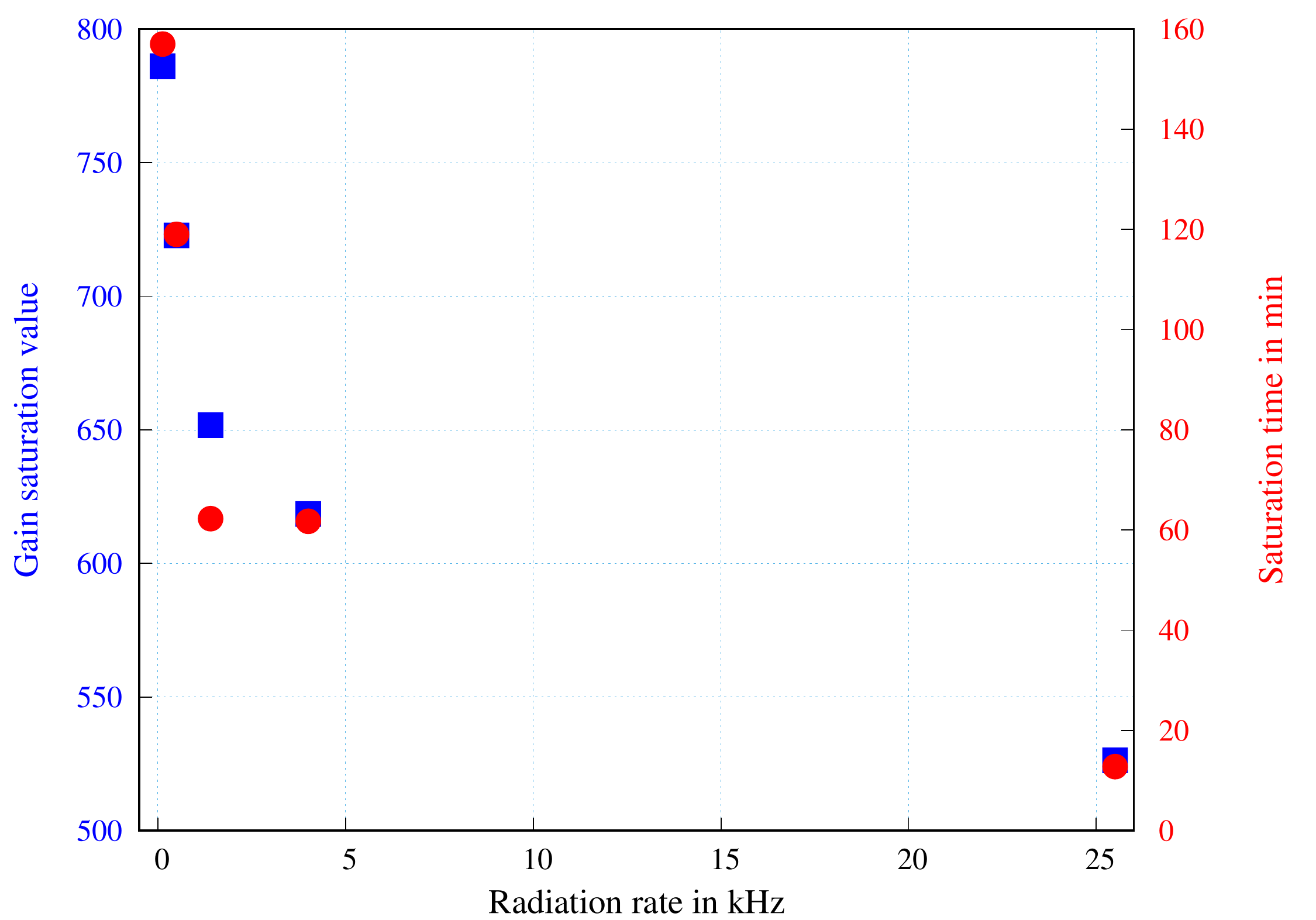}
\caption{\label{fig7b} Gain saturation value and saturation time for various radiation rates.}
\end{figure}

The value of gain is found to increase with time, for all rates. However, the functional dependence of gain with time changes for varying radiation rates because the cumulative effect of both radiation and dielectric charging-up determines the final form of the function. As the rate increases the radiation charging-up increases causing the gain to saturate at lower values (detailed discussion in section \ref{S6}). This fact is further illustrated in figure~\ref{fig7b}. Here, the variation of the steady value of the gain has been plotted as a function of the radiation rate, from the same dataset. It is seen that the final steady-state value of gain reduces as the radiation rate increases. The reduction is very pronounced at lower values of radiation rate, while for higher values of radiation rate, the variation in steady-state gain value becomes small. It is also observed that the time taken to attain the steady-state value decreases as the radiation rate increases. The derivative of the fit function in principle becomes zero when it saturates. Here the gain saturation value and its corresponding time have been obtained from the point where the derivative value becomes 0.1.

\begin{figure}[htbp]
\centering
\includegraphics[width= 0.65\linewidth,keepaspectratio]{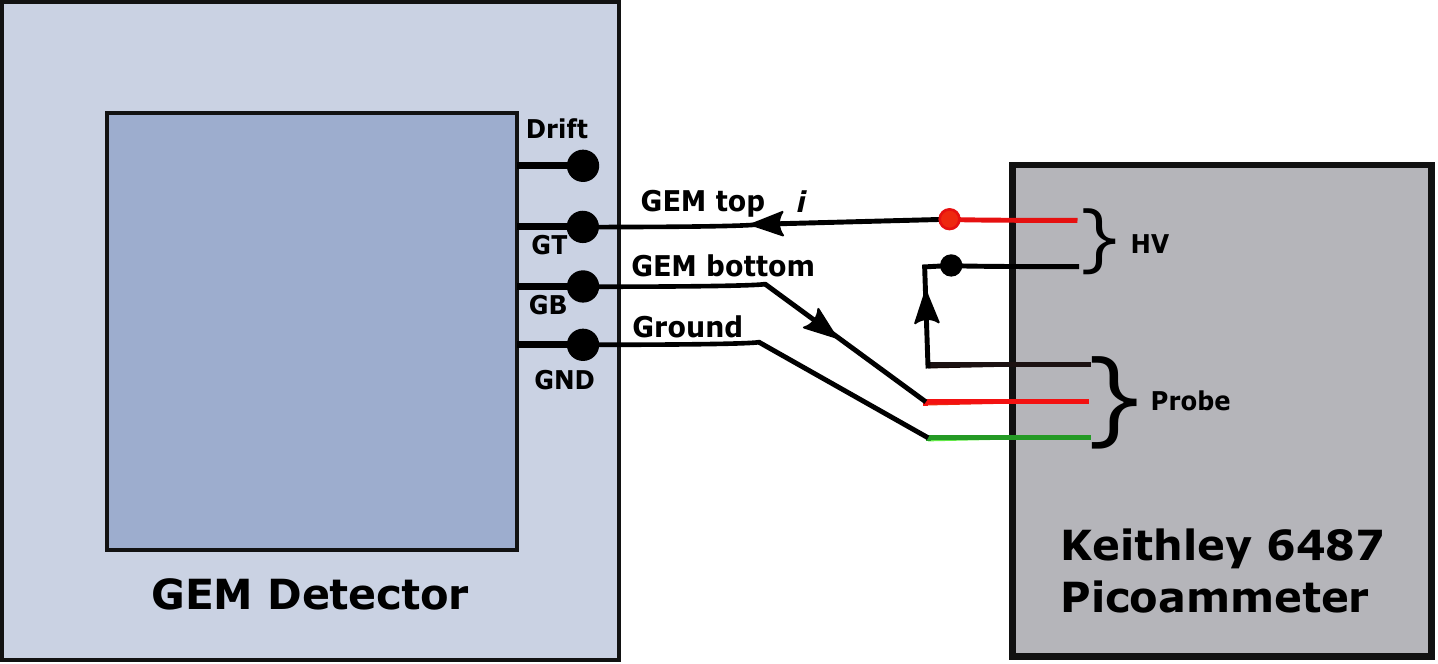}
\caption{\label{fig7c} Schematic diagram showing the connections while measuring current with Keithley 6487 picoammeter.}
\end{figure}

\begin{figure}[htbp]
\centering
\includegraphics[width= 0.65\linewidth,keepaspectratio]{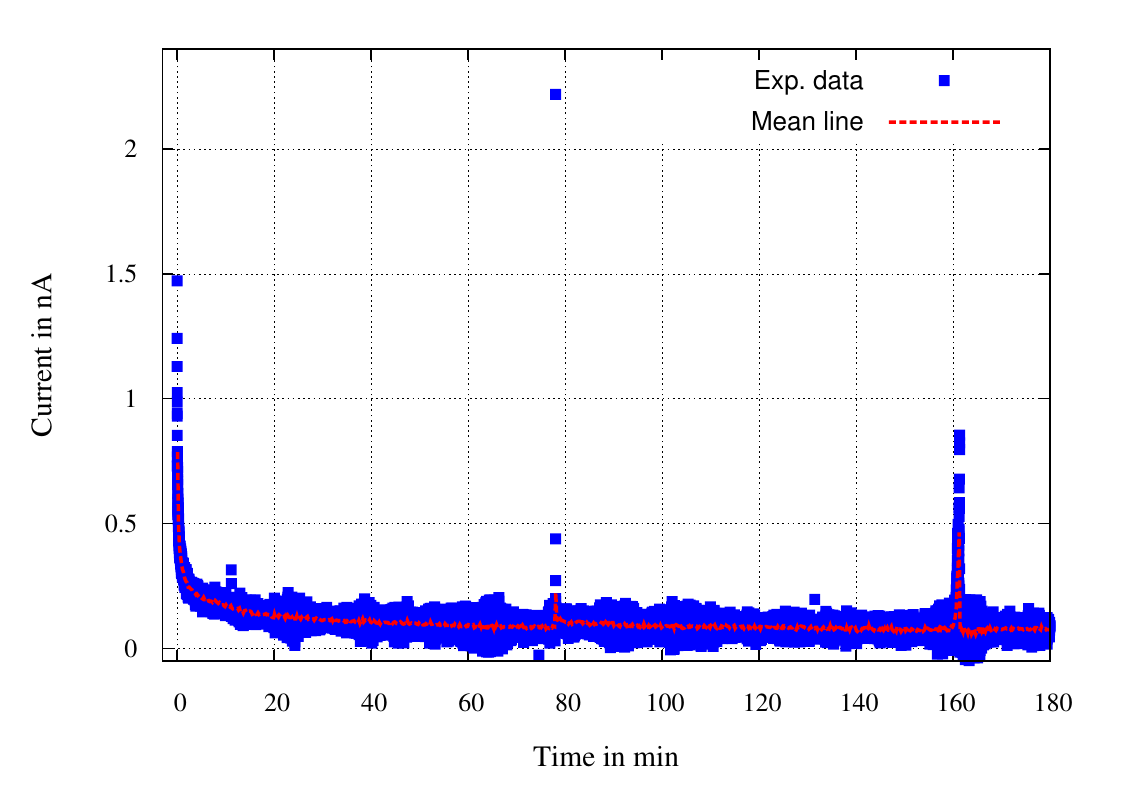}
\caption{\label{fig7d} Effect of space-charge polarization on current after applying 500V through Keithley 6487 picoammeter.}
\end{figure}

The polarization of dielectric in GEM has been further verified by an experiment in which the GEM foil have been biased using Keithley 6487 picoammeter \cite{keithley} to 500V and the current have been measured with time to see the polarization effects as shown in figure~\ref{fig7c}. The current reduces with time and finally saturates once the dielectric is completely polarized as shown in figure~\ref{fig7d} as expected from earlier works on polyimide films \cite{Sessler}.

The fluctuation in current is high in Keithley 6487 picoammeter and some peaks are also visible which can be spark or more likely the electrical noise like switching ON/OFF of Airconditionor's Compressor. The Ar-CO$_2$ gas has been flowing through the detector while carrying out this experiment.
The results obtained in figure~\ref{fig7d} and \ref{fig7} are in good agreement with the previous work \cite{Jonuz} and confirms our claim of space-charge polarization of polyimide due to application of the high electric field.

\section{Radiation charging-up}
\label{S6}
Radiation charging-up takes place when a GEM foil is irradiated with a radiation source. After multiplication inside the GEM holes, the electrons and ions move towards their respective anode and cathode planes. However, during the process of transport that involves transverse and longitudinal diffusion, some of these charges get stuck on the dielectric surface that can lead to a change in the local electric field. By distorting the local electric field, these attached charges repel further attachment of similar charges in the same location. This distorted electric field also starts to affect the overall gain of the detector. Finally, an equilibrium is expected to be achieved in which the rate at which new charges sticks to the dielectric, possibly get attached to gas atoms, or removed by the nearby copper electrodes, production of new charges through amplification in the local modified electrostatic configuration, all balance each other.

In order to ensure that the charging-up phenomenon studied has been solely due to irradiation and not dielectric polarization, the following has been maintained to measure the effect of radiation charging-up:
\begin{itemize}
\item The detector has been kept at its respective potential values for days before irradiation.
\end{itemize}

\begin{figure}[htbp]
\centering
\includegraphics[width= 0.65\linewidth,keepaspectratio]{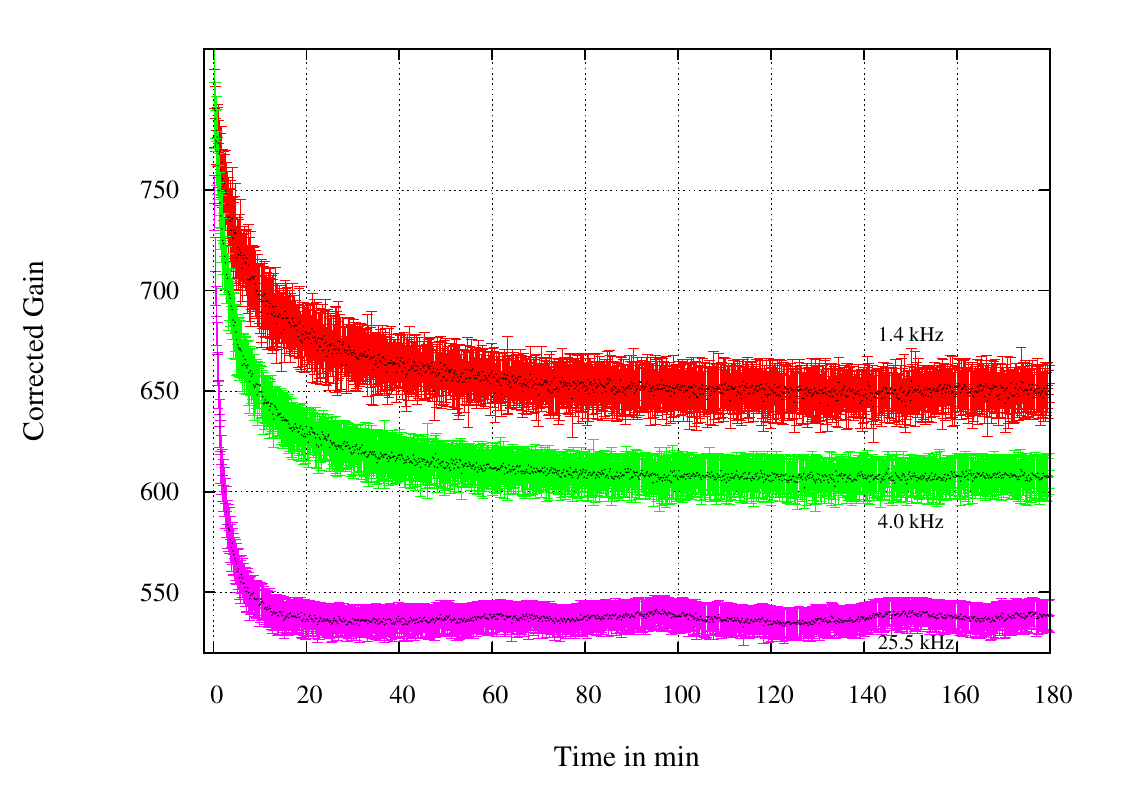}
\caption{\label{fig8} Radiation charging-up of dielectric with different radiation rates with configuration \ref{v}.}
\end{figure}

The collimator as discussed in section \ref{S2} has been used to vary the radiation equivalent rate from 1.4 to 25.5 kHz. As rate increases the rate of charges getting collected on the dielectric surface also increases. Thus, it is expected that the increase in the rate of radiation accelerates the radiation charging-up phenomenon which can be seen in figure~\ref{fig8}. These rate variation measurements were carried out with an applied potential of $\Delta$V$_{GEM} = 508.35~V$.

\begin{figure}[htbp]
\centering
\includegraphics[width= 0.65\linewidth,keepaspectratio]{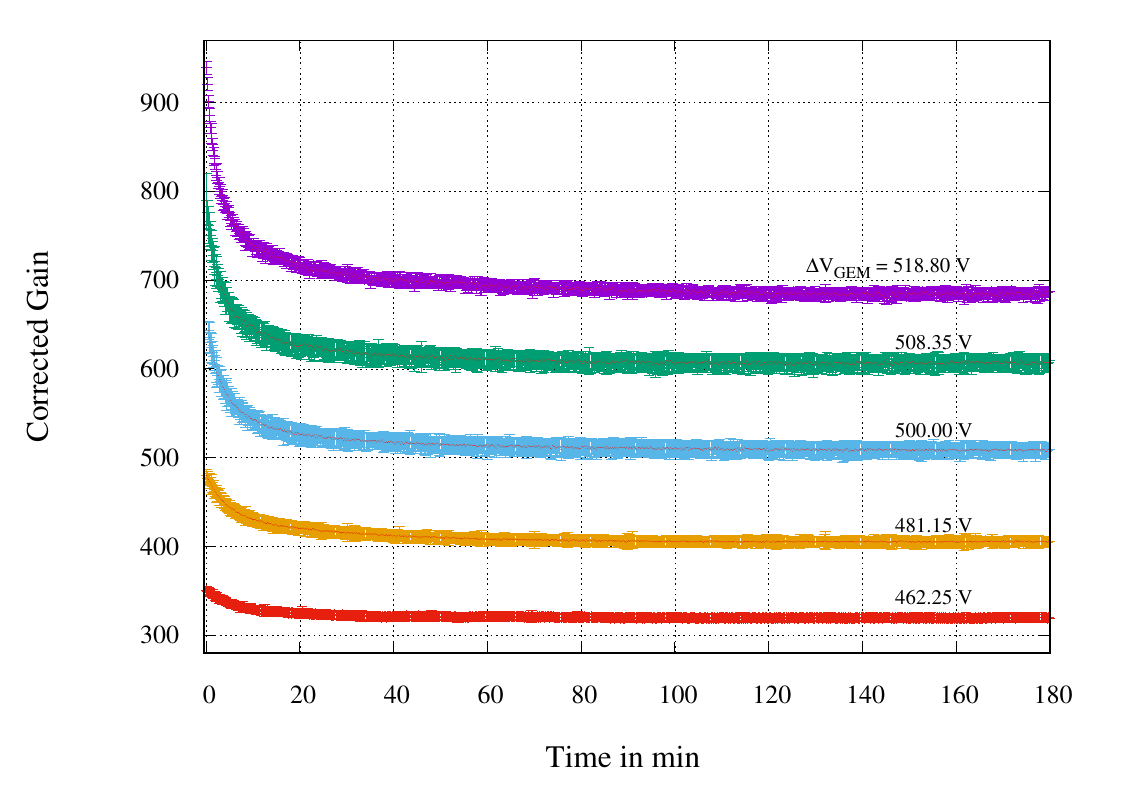}
\caption{\label{fig10} Radiation charging-up at different GEM voltages with configuration \ref{vi}.}
\end{figure}

\begin{figure}[htbp]
\centering
\includegraphics[width= 0.65\linewidth,keepaspectratio]{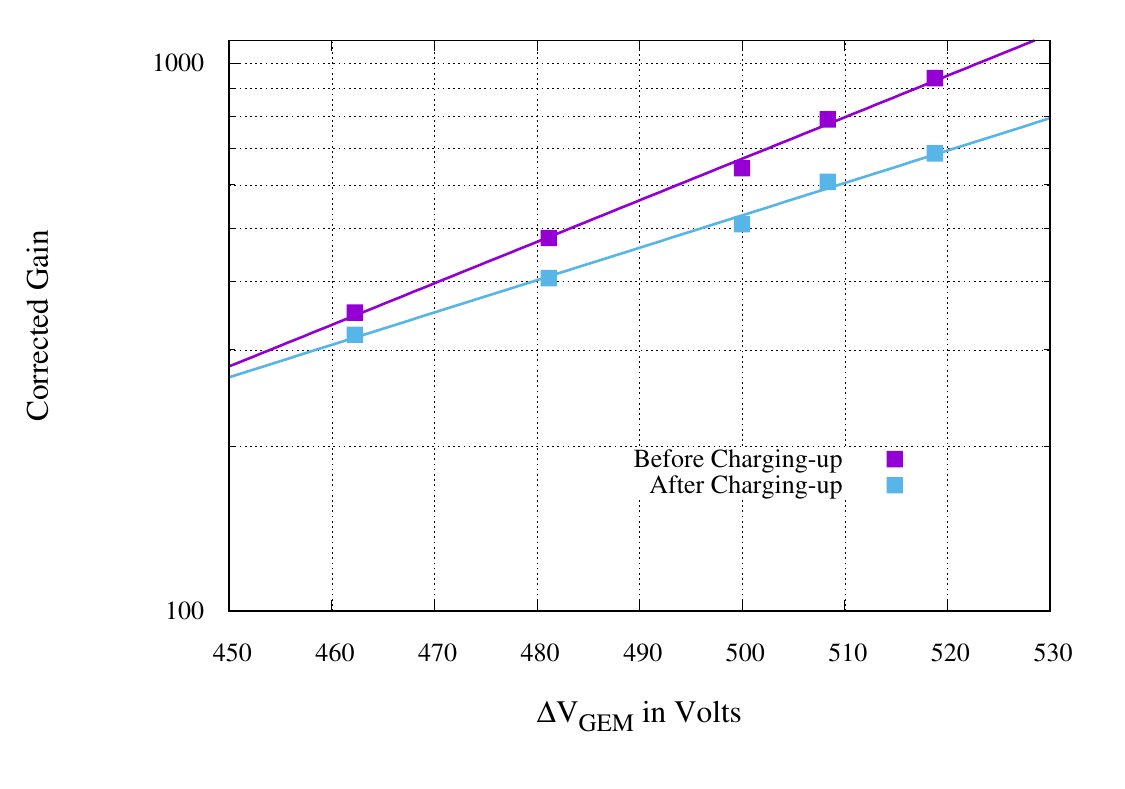}
\caption{\label{fig10b} Reduction in gain due to radiation charging-up at different GEM voltages with configuration \ref{vi}. The error is very small and within the markers.}
\end{figure}

In figure~\ref{fig10}, the effect of the application of different $\Delta$V$_{GEM}$ is studied for a fixed radiation rate. It can be seen that the steady-state value of the gain increases with an increase of $\Delta$V$_{GEM}$. This observation can be explained as follows: as the field inside the GEM foil is increased by raising the potential across the GEM foil ($\Delta$V$_{GEM}$), it causes the gain to increase by producing more number of electrons inside the GEM hole. The charging-up effect reduces the gain, but, the overall gain remains high at the higher field. The reduction in gain due to radiation charging-up caused by 4 kHz source is shown in figure~\ref{fig10b} using the same dataset.

\section{Radiation charging-down}
\label{S7}
\begin{figure}[htbp]
\centering
\includegraphics[width= 0.65\linewidth,keepaspectratio]{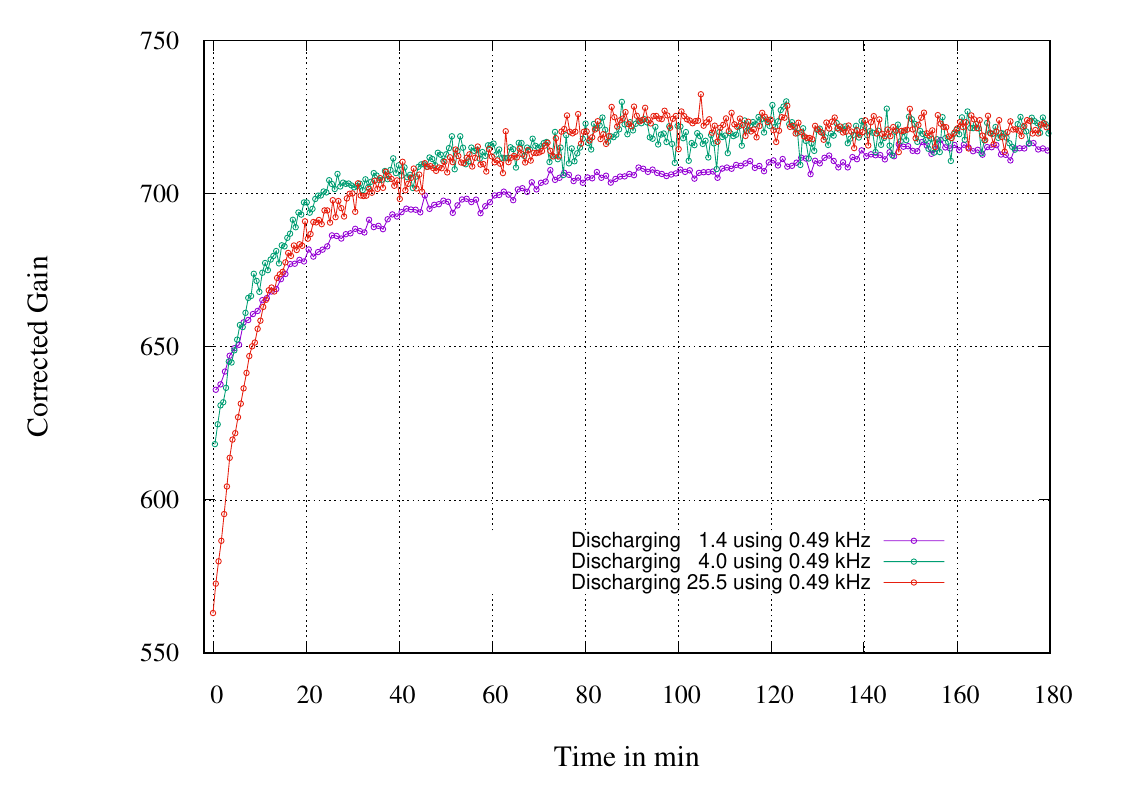}
\caption{\label{fig9} Charging-down of dielectric using 0.49 kHz after charging-up with 1.4, 4.0, 25.5 kHz with configuration \ref{vii}. To see the curves clearly error bars were not used. }
\end{figure}

To study radiation charging-down of an already charged-up GEM dielectric, a test probe of comparatively low radiation rate has been employed for gain measurement during the charging-down phase of the detector. The collimator used for the test probe is of 2mm diameter with a total radiation rate of 490 Hz. This radiation is large enough to give us the energy spectra within a small interval of time $\Delta$t, to see the charging-down curve and small enough to ensure comparatively low radiation charging-up.
With the removal of the high rate source and use of a weak test probe, the trapped charges decrease gradually with very few new charges created to replace them. The rate of this process depends on the attachment coefficient of the gas mixture, conductivity and mobility of charges in the dielectric, as well as on the number density of trapped charge in the material \cite{azmoun2006,correira2014} apart from other environmental factors like temperature, pressure and humidity. As the amount of charge trapped decreases, the field, in turn, comes to its original value leading to an increase in the gain. This is displayed in figure~\ref{fig9} which shows the effect of charging-down on the gain, after the GEM was charged up by applying radiation.

The experiment has been carried out using the following steps:
\begin{itemize}
\item The detector has been charged-up with high radiation rate (measuring radiation charging-up).
\item Once the gain saturates implying that the detector has completely charged-up, the high rate source has been replaced by the test probe: 490 Hz radiation rate with 2 mm collimator.
\item Measurements related to charging down have been performed using the test probe.
\item Steps mentioned above have been repeated with various radiation rates for charging-up and charging-down measurement.
\item To be on the safe side, the high rate source was placed for at least six hours before replacing it with the test probe ensuring complete charging-up of GEM foil.
\item For the entire duration of these measurements related to figures~\ref{fig8} and \ref{fig9}, the potential applied to the detector was kept fixed at $\Delta$V$_{GEM} = 508.35~V$.
\end{itemize}

\section{Radiation charging-up with two different sources}
\begin{figure}[htbp]
\centering
\includegraphics[width= 0.65\linewidth,keepaspectratio]{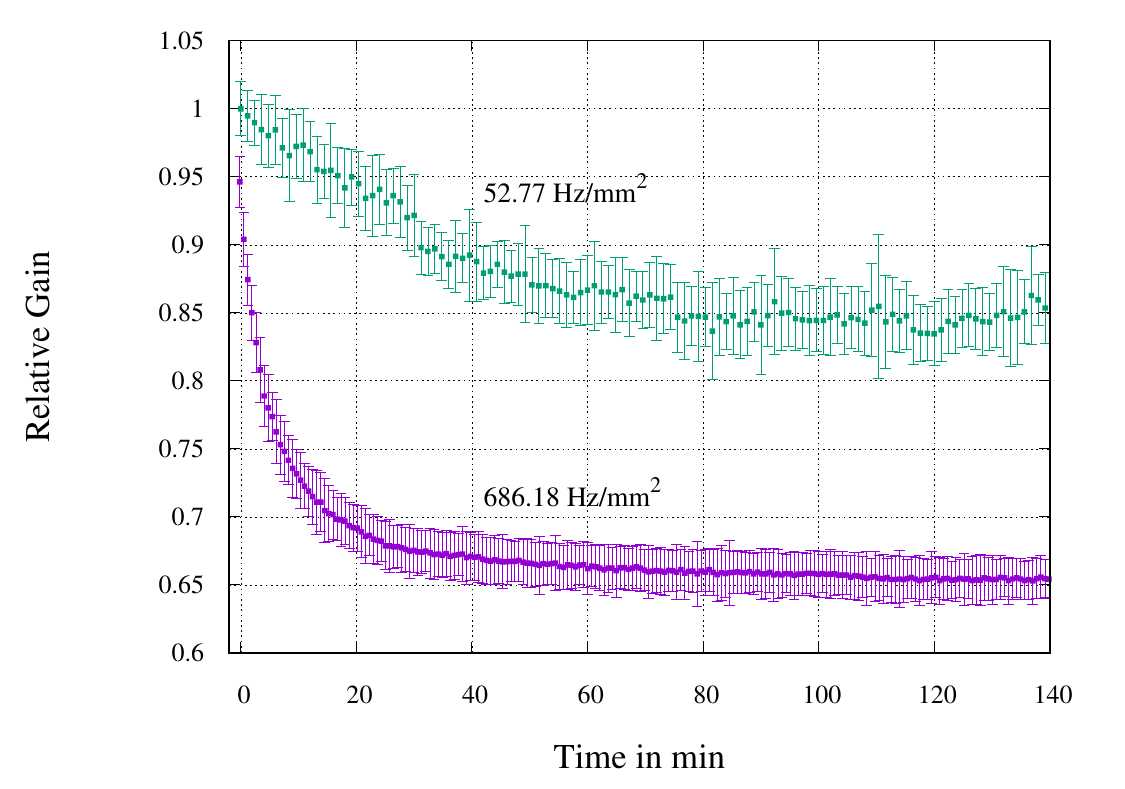}
\caption{\label{fig12}Radiation charging-up with two different well-collimated sources having different rates.}
\end{figure}

A separate set of measurements has been carried out to verify the results discussed in section \ref{S6}. Previously the rate have been controlled by changing the aperture size of the collimator (increasing both exposure area, as well as the radiation rate, as shown in figure~\ref{fig2e}). Here, two sources of different rates have been used. The gas mixture used is Ar-CO$_2$ (90-10 $\pm 1\%$) with 1 kV/cm and 2.5 kV/cm induction and drift fields respectively, $\Delta$V$_{GEM} =$ 430 V. As before the detector was kept at its respective potential for days before measurements, the irradiation starts at time t=0 sec. As shown in figure~\ref{fig12} similar effects of radiation charging-up were observed with the high rate source causing more reduction of gain than the lower one. A collimator disk of thickness 12 mm with an aperture size of 4mm have been used for both the source to ensure the radiation beam is parallel unlike the thin collimator disk used in section \ref{S6}.

\section{Summary and Conclusions}
The effect of both radiation and polarization charging-up on the gain of a single GEM has been studied using experimental techniques. The amount of polarization has been controlled by the voltage applied to the GEM foil, while the rate of radiation entering the detector has been controlled by the use of collimators of various sizes. Individual and combined effects of these two parameters have been studied. The charging-up due to polarization of dielectric is found to increase the gain whereas, the gain is found to decrease due to radiation charging-up which is in agreement with the experimental studies in \cite{azmoun2006},\cite{croci2009} and the numerical simulations in \cite{alfonsi2012simulation}. When the rate of irradiation is high, the time taken for radiation charging-up is less to attain gain equilibrium.
None of these effects are found to be permanent and the detector comes back to its normal state once the biasing and radiation source are removed. In our experimental setup, it has taken a couple of hours to revert to normal.

\acknowledgments
The authors would like to acknowledge Mr. Shaibal Saha and Mr. Pradipta K. Das for their technical help in assembling of single GEM and designing of the potential divider circuit. We also acknowledge Mr. Anil Kumar for his valuable discussion, suggestions in data analysis, and programming of Arduino Uno board for temperature and pressure measurements. The authors are grateful to Mr. Subendu Das for his contribution in programing the Field-Programmable Gate Array (FPGA) board used to control the CAEN N471 power supply using TTL signal which was of great help in dielectric charging-up measurements. This work has partly been performed in the framework of the RD51 Collaboration. We wish to acknowledge the members of the RD51 Collaboration for their help and suggestions.

\end{document}